\documentclass[12pt,a4paper]{article}
\usepackage{epsfig}
\begin{document}
\title{Photon-induced production of the mirror quarks
from the $LHT$ model at the $LHC$}
\author{Chong-Xing Yue, Jin-Yan Liu, Li Ding, Wei Liu, Wei Ma  \\
{\small Department of Physics, Liaoning  Normal University, Dalian
116029, P. R. China}
\thanks{E-mail:cxyue@lnnu.edu.cn}}
\date{\today}

\maketitle
\begin{abstract}

\vspace{1cm}

The photon-induced processes at the $LHC$ provide clean experimental
conditions due to absence of the proton remnants, which might
produce complementary and interesting results for tests of the
standard model and for searching of new physics. In the context of
the littlest $Higgs$ model with T-parity, we consider the
photon-induced production of the mirror quarks at the $LHC$. The
cross sections for various production channels are calculated and a
simply phenomenology analysis is performed by assuming leptonic
decays.

 \vspace{2.0cm} \noindent

\end{abstract}
\newpage
\noindent{\bf 1. Introduction}

It is well known that the equivalent photon approximation ($EPA$)
can be successfully to describe most of the processes involving
photon exchange [1]. A significant fraction of $pp$ collisions at
the $LHC$ will involve quasi-real photon interactions occurring at
energies well beyond the electroweak energy scale [2]. Therefore,
the $LHC$ can be considered as a high-energy photon-photon or
photon-proton collider, which is paid significant attention recently
[3].

The photon-induced processes offer a rich and exciting field of
research at the $LHC$ [3]. In general, the exclusive two-photon
production, $pp \rightarrow p \gamma \gamma p \rightarrow p X p$,
provides clean experimental conditions and well defined final
states, which can be selected and precisely reconstructed. Moreover,
 for the dedicated very forward detectors ($VFDs$),
 detection of the two final state protons, scattered at almost
zero-degree angle, provides another striking signature, effective
also at high luminosity and with large event pile-up [2,4]. Thus,
the two-photon production of the charged particle pairs offers
interesting potential for signals of new physics beyond the standard
model ($SM$) at the $LHC$. It has been shown that detection of the
supersymmetric charginos, sleptons and the charged Higgs bosons is
very unambiguous in two-photon exclusive production, allowing for
clear interpretation [5,6]. The exclusive two-photon production of
 the boson pairs $WW$ and $ZZ$ at the $LHC$ provides an excellent way
to test the electroweak gauge boson sector [7].

The luminosity and the center-of-mass ($c.m.$) energy of
photon-proton ($\gamma p$) collisions are higher than the
photon-photon ones at the $LHC$. This offers interesting
possibilities for the study of electroweak interactions and for
searching the new physics beyond the $SM$ up to $TeV$ scale [8]. In
contrast to the photon-photon production processes, the
photoproduction processes involve topologies with hard jets in the
final state. The effect of jet algorithms and the efficiency of
event selection was taken into account using a fast simulation of a
typical multipurpose $LHC$ detector response [8]. Thus, a large
number of $pp \rightarrow p (\gamma g/q)Y \rightarrow pXY$ processes
have sizable cross sections and could be studied during the very low
and low luminosity phases of the $LHC$. So far, the cross sections
for many electroweak processes and some new physics processes with
their irreducible background processes are studied in Refs.[7,8,9].

The little $Higgs$ theory [10] is one of the interesting candidates
of the new physics beyond the $SM$. The little $Higgs$ model with
$T$-patity (called the $LHT$ model) [11] is one of the attractive
little $Higgs$ models. In this model, particles are divided into
$T$-even and $T$-odd sectors under $T$-parity. The T-even sector
consists of the $SM$ particles and a heavy top $T_{+}$, while the
T-odd sector contains heavy gauge bosons
($B_{H},Z_{H},W_{H}^{\underline{+}}$), a scalar triplet ($\Phi$),
and the so-called mirror fermions. These new particles can produce
rich phenomenology at present and in future high energy collider
experiments [12,13,14,15,16]. Reference [17] has considered
photoproduction of pairs of T-odd particles via $e^{+}e^{-}$ and
$ep$ collisions. As we know, so far, photoproduction and two-photon
production of the T-odd particles predicted by the $LHT$ model have
not been considered at the $LHC$, which is the main aim of this
paper.

In this paper, we consider the photoproduction and two-photon
production processes involved the mirror quarks at the $LHC$. We
show that most of their production cross sections are smaller than
those coming from the partonic processes for the mirror quarks at
the $LHC$. However, considering their clean experimental conditions
and well defined final states, these production processes should be
further studied. The photoproduction and two-photon production
processes for the mirror quarks might help to search possible
signatures of the $LHT$ model at the $LHC$.

In the rest of this paper, we will give our results in detail. In
section 2, we briefly review the essential features of the $LHT$
model. Photoproduction of the mirror
 quark associated with a new gauge boson and with a new scalar are
 discussed in sections $3$ and $4$, respectively. The cross sections
 for photoproduction and two-photon production of the mirror
 quark pairs are calculated in section $5$. The simple phenomenology analysis at
 the $LHC$ are also given in these sections. Finally, we summarize
 our results and given some discussions in section $6$.

\noindent{\bf 2. The essential features of the $LHT$ model}

In this section, we briefly review the essential features of the
$LHT$ model studied in Refs.[11,12], which are related to our
calculation. The $LHT$ model is based on a $SU(5)/SO(5)$ global
symmetry breaking pattern, which gives rise to fourteen
Number-Goldstone ($NG$) bosons. Four of the fourteen $NG$ bosons are
eaten by the T-odd heavy gauge bosons ($B_{H},Z_{H},W_{H}^{\pm}$)
associated with the gauge symmetry breaking $[SU(2) \times U(1)]_{1}
\times [SU(2) \times U(1)]_{2} \rightarrow
 SU(2)_{L} \times U(1)_{Y}$ at the scale $f$. The remaining $NG$
bosons decompose into a T-even doublet $H$, considered to be the
$SM$ $Higgs$ doublet, and a complex T-odd $SU(2)$ triplet $\Phi$
\begin{equation}
 H=\left(
 \begin{array}{c}-i\frac{\pi^{+}}{\sqrt{2}}\\\frac{\nu+h+i\pi^{0}}{2}
 \end{array}\right),
 \hspace{1.5cm} \Phi=\left(
 \begin{array}{cc}-i\phi^{++}&-i\frac{\phi^{+}}{\sqrt{2}}\\
 -i\frac{\phi^{+}}{\sqrt{2}}&\frac{-i\phi^{0}+\phi^{P}}{\sqrt{2}}\end{array}\right).
\end{equation}
Here $h$ is the physical Higgs field and $\nu = 246 GeV$ is the
electroweak scale. The $\pi^{0,\pm}$ are absorbed by the $SM$ gauge
bosons $W^{\pm}$ and $Z$ after electroweak symmetry breaking
($EWSB$). There is an relation between the T-odd triplet and the
T-even $Higgs$ boson masses, which is approximately expressed as
[12]
\begin{equation}
     M_{\Phi}=\frac{\sqrt{2}m_{H}}{\nu} f,
\end{equation}
where $f$ is the scale parameter of the gauge symmetry breaking of
the $LHT$ model. At the leading order, the components $\phi^{+}$,
$\phi^{-}$, $\phi^{0}$, and $\phi^{p}$ of the triplet scalar $\Phi$
have same mass, i.e.
$M_{\phi^{+}}=M_{\phi^{-}}=M_{\phi^{0}}=M_{\phi^{p}}=M_{\Phi}$ .

After taking into account $EWSB$, at the order of $\nu^{2}/f^{2}$,
the masses of the $T$-odd set of the $SU(2)\times U(1)$ gauge bosons
are given by
\begin{eqnarray}
 M_{B_{H}}=\frac{g'f}{\sqrt{5}}[1-\frac{5\nu^{2}}{8f^{2}}],\hspace{0.5cm}
 M_{Z_{H}}\approx
  M_{W_{H}}=gf[1-\frac{\nu^{2}}{8f^{2}}].
\end{eqnarray}
Where $g'$ and $g$ are the $SM$ $U(1)_{Y}$ and $SU(2)_{L}$ gauge
coupling constants, respectively. Because of the smallness of $g'$,
the gauge boson $B_{H}$ is the lightest T-odd particle, which is
stable, electrically neutral, and weakly interacting particle. Thus,
it can be seen as an attractive dark matter candidate [12].
Certainly, if the T-parity is violated by anomalies, the lightest
T-odd gauge boson $B_{H}$ can decay into the $SM$ gauge boson pairs
$WW$ and $ZZ$ [18,19].

To avoid severe constraints and simultaneously implement $T$ parity,
one needs to double the SM fermion doublet spectrum [11, 12]. The
$T$-even combination is associated with the SM $SU(2)_{L}$ doublet,
while the $T$-odd combination is its $T$-parity partner, which are
called the mirror fermions. Assuming  there is flavor universal and
diagonal Yukawa coupling $k$, the mirror quarks for different
generations will be degenerate in mass, and the masses of the up-
and down-type mirror fermions can be written as [13]
\begin{equation}
M_{U_{H}}\approx\sqrt{2}kf(1-\frac{\nu^{2}}{8f^{2}}),\hspace{0.5cm}
M_{D_{H}}\approx\sqrt{2}kf.
\end{equation}
Being $f\geq$500GeV, it is clear from Eq.(4) that there is
$M_{U_{H}} \approx M_{D_{H}} $. Thus, we can assume the mirror
quarks degenerating in mass and take $M_{Q_{H}}=M=\sqrt{2}kf$. This
means that the mirror quarks have no contributions to the flavor
changing processes, which is the minimal flavor violation ($MFV$)
limit of the $LHT$ model [20]. In this paper, we will focus our
attention on production of the first and second generation mirror
quarks via photon interactions at the $LHC$ and assume that the
value of the coupling constant $k$ is in the range of $ 0.5\sim
1.5$.

The mirror quarks can couple to ordinary quarks mediated by the
T-odd gauge bosons and at higher order by the scalar triple $\Phi$,
which are parameterized by two $CKM$-like unitary mixing matrices
$V_{Hu}$ and $V_{Hd}$. They satisfy $V_{Hu}^{+}V_{Hd}=V_{CKM}$, in
which the $CKM$ matrix is defined through flavor mixing in the
down-type quark sector [15,20]. The coupling expressions, which are
related our calculation, are given in Ref.[15]. Using these Feynman
rules, we can calculate the production cross sections of the mirror
 quarks via $\gamma p$ and $\gamma \gamma$ collisions at the $LHC$.

 From the above discussions, we can see that the cross sections for
 photoproduction and two-photon production of the mirror quarks are
 generally dependent on the model parameters $f$, $k$, $(V_{Hu})_{ij}$ and
 $(V_{Hd})_{ij}$. The matrix $V_{Hd}$ can be parameterized in terms
 of three mixing angles and three phases, which can be probed by the
 flavor changing neutral current ($FCNC$) processes in K and B meson
 systems, as discussed in detail in Refs.[15,20]. To avoid any
 additional free parameters introduced and to simply our
 calculation, we take the structure of the mixing matrix $V_{Hd}$ as
$V_{Hd}=I$, which means $V_{Hu}=V_{CKM}^{+}$ and the mirror quarks have no impact
on the $FCNC$ processes.

\noindent{\bf 3. Photoproduction of the mirror quark associated with
a T-odd gauge boson}

From the above discussions, we can see that the mirror quark can be
produced associated with a T-odd gauge boson via the subprocess $q
\gamma \rightarrow Q_{H}B_{H}$ ($Z_{H}$ or $W_{H}$) at the $LHC$.
The relevant Feynman diagrams are depicted in Fig.1.
\begin{figure}[htb]
\begin{center}
\epsfig{file=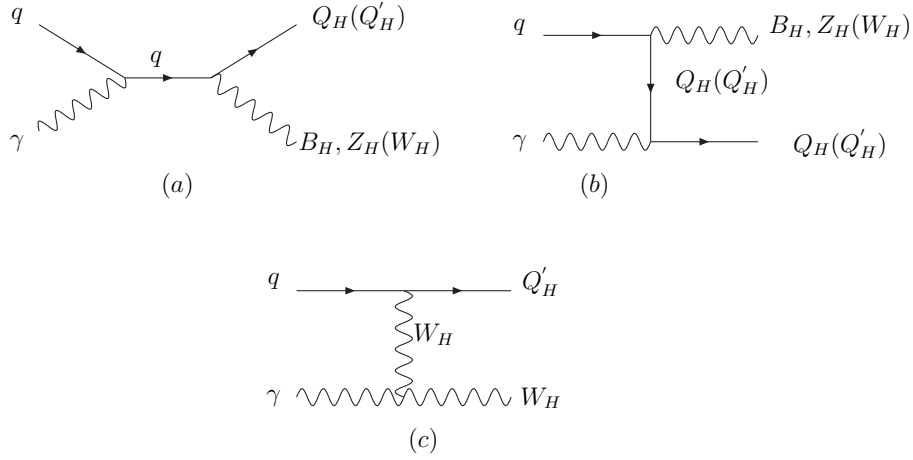,scale=0.8}
 \caption{Feynman diagrams for photoproduction of the mirror quark associated
   with \hspace*{2.0cm}a new gauge boson.}
 \label{ee}
\end{center}
\end{figure}
\vspace*{-0.8cm}

Using the relevant Feynman rules given in Ref.[15], the
corresponding scattering invariant amplitudes can be written as
\begin{eqnarray}
M_{B}^{ij}= -\frac{i}{10} \frac{e^{2}}{C_{W}} q^{i}
V_{ij}\bar{u}(P_{Q})[\frac{\not\!\!
\epsilon_{2}P_{L}(\not\!\!P_{\gamma}+\not\!\!P_{q})\not\!\!\epsilon_{1}}
{\hat{s}}+\frac{\not\!\!
\epsilon_{1}(\not\!\!P_{B}-\not\!\!P_{q}+M)\not\!\!\epsilon_{2}P_{L}}
{\widehat{t}_{B}-M^{2}}]u(P_{q}),
\end{eqnarray}

\begin{eqnarray}
M_{Z}^{ij}= \pm \frac{i}{2} \frac{e^{2}}{S_{W}} q^{i}
V_{ij}\bar{u}(P_{Q})[\frac{\not\!\!
\epsilon_{2}P_{L}(\not\!\!P_{\gamma}+\not\!\!P_{q})\not\!\!\epsilon_{1}}
{\hat{s}}+\frac{\not\!\!
\epsilon_{1}(\not\!\!P_{Z}-\not\!\!P_{q}+M)\not\!\!\epsilon_{2}P_{L}}
{\widehat{t}_{Z}-M^{2}}]u(P_{q}),
\end{eqnarray}

\begin{eqnarray}
M_{W}^{ij}= \frac{i}{\sqrt{2}} \frac{e^{2}}{S_{W}}
V_{ij}\bar{u}(P_{Q})\{\frac{\not\!\!
\epsilon_{2}P_{L}(\not\!\!P_{\gamma}+\not\!\!P_{q})\not\!\!\epsilon_{1}
q^{i}}{\hat{s}}+\frac{\not\!\!
\epsilon_{1}(\not\!\!P_{W}-\not\!\!P_{q}+M)\not\!\!\epsilon_{2}P_{L}
q^{i}}{\widehat{t}_{W}^{
1}-M^{2}}\\\nonumber+
\frac{\not\!\!\epsilon_{2}P_{L}[g^{\mu\rho}P_{\gamma}^{\sigma}+
g^{\mu\sigma}(P_{Q}-P_{q}-P_{r})^{\rho}+g^{\rho\sigma}
(P_{q}-P_{Q})^{\mu}]}{\widehat{t}_{W}^{ 2}-M_{W_{H}}^{2}}\}u(P_{q}).
\end{eqnarray}
Where $\hat{s}=(P_{\gamma}+P_{q})^{2}$,
$\widehat{t}_{B}=(P_{B}-P_{q})^{2}$,
$\widehat{t}_{Z}=(P_{Z}-P_{q})^{2}$,
$\widehat{t}_{W}^{1}=(P_{W}-P_{q})^{2}$, and
$\widehat{t}_{W}^{2}=(P_{Q}-P_{q})^{2}$. $i$ represents the $SM$
light quark $u$, $c$, $d$, or $s$, and $q^{i}$ represents the
corresponding electric charge. $j$ is the family indexes for the
mirror quarks and $M$ is the mass of the mirror quark.
$S_{W}=sin\theta_{W}$, $C_{W}=cos\theta_{W}$, and $\theta_{W}$ is
the Weinberg angle. $\varepsilon_{1}$  is the polarization vector of
the photon $\gamma$ and $\varepsilon_{2}$ is the polarization vector
of the gauge boson $Z_{H}$, $B_{H}$, or $W_{H}$.
$P_{L}=({1-\gamma_{5}} )/ {2}$ is the left-handed projection
operator. For the up-type quark $u$ or $c$, the $CKM$-like matrix
element $V_{ij}$ is $(V_{Hu})_{ij}$, while for the down-type quark
$d$, $s$, or $b$, $V_{ij}$ is $(V_{Hd})_{ij}$. In Eq.(6), $ " \pm "$
represent the up- and down-type quarks, respectively.

\begin{figure}[htb]
\begin{center}
\epsfig{file=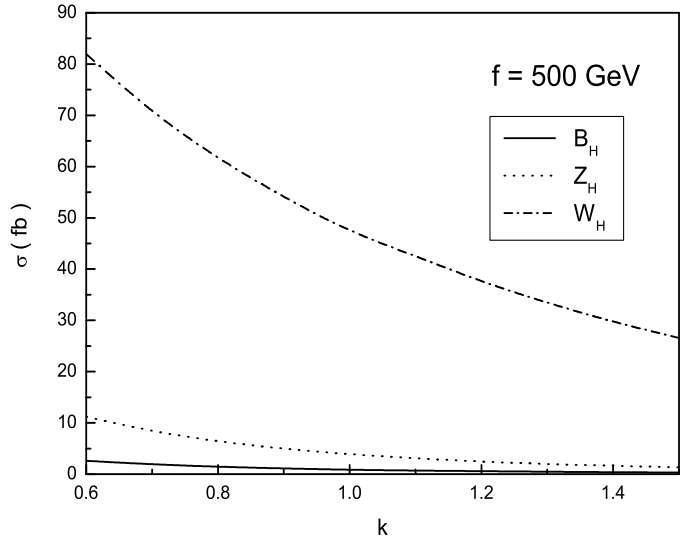,width=285pt,height=240pt}
 \vspace{-0.9cm}
 \caption{The cross sections for the photoproduction of the mirror
 quarks associated with \hspace*{1.8cm}
 the new gauge bosons ($B_{H}$, $Z_{H}$, $W_{H}$) as functions of
 the coupling constant $k$
\hspace*{1.8cm}
for the scale parameter $f=500GeV$.}
 \label{ee}
\end{center}
\end{figure}

After calculating the cross section
$\widehat{\sigma}_{G}^{ij}(\widehat{s})$ ($G$ = $Z_{H}$, $B_{H}$, or
$W_{H}$) for the subprocess $q\gamma \rightarrow Q_{H}B_{H}$($Z_{H}$
or $W_{H}$), the effective cross section $\sigma_{G}$ at the $LHC$
can be obtained by folding $\widehat{\sigma}_{G}^{ij}(\widehat{s})$
with the parton distribution functions ($PDFs$)
\begin{eqnarray}
\sigma_{G}=\sum_{i,
j}\int^{1}_{x_{min}}\int^{\tau_{max}}_{\tau_{min}} dx d \tau
f_{q_{i}/p}(x, \mu) f_{\gamma /p}(\tau)
\hat{\sigma_{G}}^{ij}(\hat{s})
\end{eqnarray}
with $x_{min}=(M_Q+M_{G})^{2}/S$, $\tau_{min}=(M_{Q}+M_{G})^{2}/Sx
$, $\tau_{max}=(1-m/\sqrt{S})^{2}$ and $\hat{s}=x\tau S$, in which
the $c.m.$ energy $\sqrt{S}$ is taken as $14TeV$ for the $LHC$. $m$
is the proton mass. In our numerical calculation, we will use
$CTEQ6L$ $PDF$[21] for the quark distribution $f_{q_{i}/p}(x,
\mu_{F})$ and assume that the factorization scale $\mu_{F}$ is of
order $\sqrt{\widehat{s}}$. The photon distribution function
$f_{\gamma /p}(\tau)$ is assumed that it only is the elastic
components of the equivalent photon distribution of the proton,
which has been extensively studied in Refs. [1, 22, 23].

\vspace{0.8cm}
\begin{figure}[htb]
\begin{center}
\vspace{-0.5cm}
 \epsfig{file=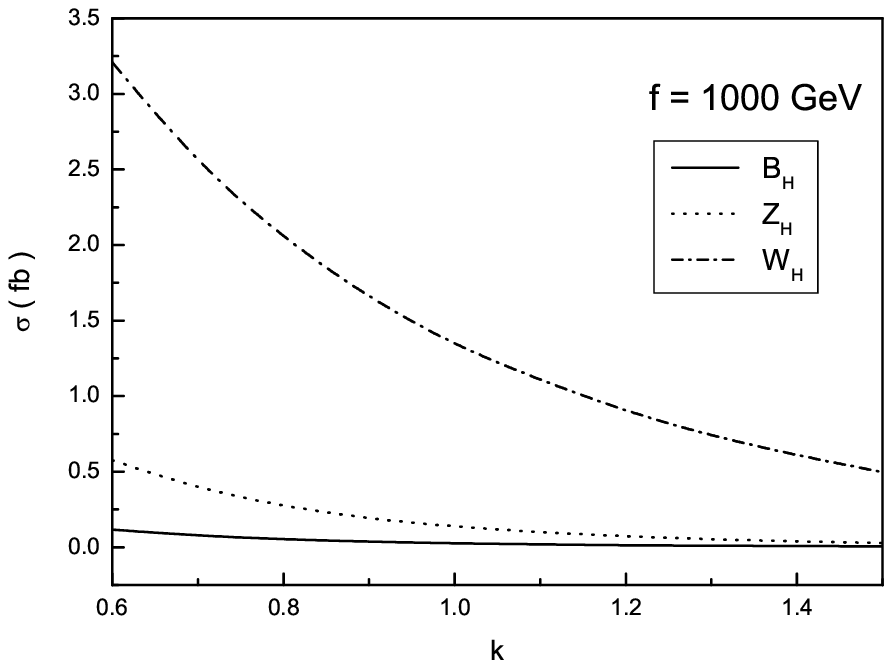,width=200pt,height=165pt}
\put(-115,-10){ (a)}\put(115,-10){ (b)}
 \hspace{0cm}\vspace{-0.25cm}
\epsfig{file=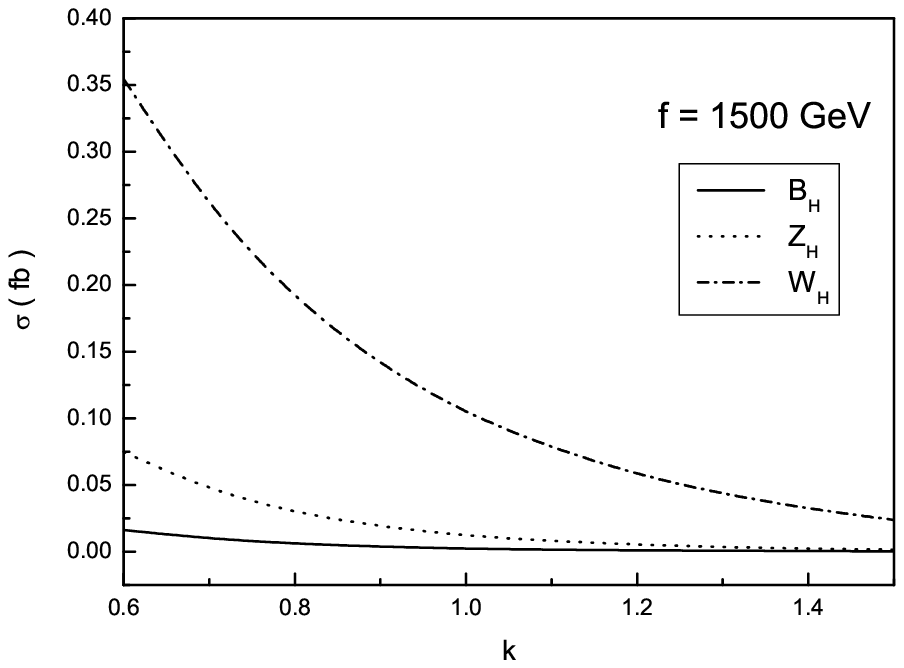,width=200pt,height=165pt} \hspace{-0.5cm}
 \hspace{10cm}\vspace{-1cm}
\vspace{0.5cm}
 \caption{(a) Same as Fig.2 but for $f=1TeV$. (b) Same as Fig.2 but
for $f=1.5TeV$.}
 \label{ee}
\end{center}
\end{figure}

\vspace{-0.6cm}

 In Fig.2, Fig.3(a), and
Fig.3(b) we plot the cross sections for photoproduction of the
mirror quark associated with the new gauge bosons as functions of
the parameter $k$ for three values of the scale parameter $f$. In
these three figures we have taken the values of the $CKM$ matrix
elements $(V_{CKM})_{ij}$ given in Ref.[24], in which $V_{CKM}$ is
constructed based on the parameterization [25]. One can see from
these figures that, although there is the relation $M_{B_{H}}<
M_{Z_{H}}\simeq M_{W_{H}}$, the cross section $\sigma_{W_{H}}$ is
most large and the cross section $\sigma_{B_{H}}$ is smaller than
$\sigma_{W_{H}}$ or $\sigma_{Z_{H}}$. This is because the coupling
constants of the new gauge boson $B_{H}$ to the mirror quarks and
ordinary quarks are smaller than those for the new gauge bosons
$Z_{H}$ and $W_{H}$. Furthermore, for the subprocess $q \gamma
\rightarrow Q_{H}^{'}W_{H}$, there is an extra Feynman diagram as
shown in Fig.1(c) contributing to the cross section
$\sigma_{W_{H}}$. For $0.6\leq k\leq 1.4$ and $500GeV\leq f \leq
1500GeV$, the values of the effective production cross sections
$\sigma_{B_{H}}$, $\sigma_{Z_{H}}$ , and $\sigma_{W_{H}}$ are in the
ranges of $1.6\times10^{-2}fb \sim 2.6fb$, $7.5\times10^{-2}fb \sim
11fb$, $0.36\times10^{-2}fb \sim 82fb$, respectively.

For $k>0.5$, the mirror quark is heavy enough to decay into $T$-odd
gauge boson plus an ordinary fermion. The branching ratios of the
possible two-body modes of the mirror quarks $U_{H}$ and $D_{H}$ are
discussed in Refs.[13,26]. The different chain decays of the mirror
quark can given different experimental signatures at the $LHC$,
which have been extensively studied. The up- and down- type mirror
quarks mainly decay into $d W_{H}^{+}$ and $uW_{H}^{-}$,
respectively. For $0.6\leq k\leq 1.4$ and $f=1TeV$, the values of
the branching ratios $Br(U_{H}\rightarrow d W_{H}^{+})$ and
$Br(D_{H}\rightarrow u W_{H}^{-})$ are all about $57$\% [26]. To
simply our phenomenology analysis, we only consider the signatures
induced by the decay modes $uW^{-}_{H}$ and $dW^{+}_{H}$ and take
these two decay modes as $jW_{H}$, in which $j$ indicates a
light-flavor jet $u$, $c$, $d$, or $s$. Furthermore, we will assume
that T-parity is strictly conserved and the T-odd gauge boson
$B_{H}$ can be seen as missing energy.

From the above discussions, we can see that photoproduction of the
mirror quark associated with the T-odd gauge boson at the $LHC$ can
give the following final states
 \begin{eqnarray}
 q \gamma \rightarrow Q_{H}B_{H}\rightarrow j W_{H}
 B_{H}\rightarrow j W B_{H}B_{H} \rightarrow j l \nu_{l}B_{H}B_{H},
 \end{eqnarray}
\vspace*{-1.1cm}
 \begin{eqnarray}
q \gamma \rightarrow Q_{H}Z_{H}\rightarrow j W_{H} Z_{H}\rightarrow
j W H B_{H}B_{H} \rightarrow j l \nu_{l}b\overline{b}B_{H}B_{H},
 \end{eqnarray}
\vspace*{-1.1cm}
 \begin{eqnarray}
q \gamma \rightarrow Q_{H}W_{H}^{\pm}\rightarrow j W_{H}^{\pm}
W_{H}^{\mp}\rightarrow j W^{\mp}W^{\pm} B_{H}B_{H} \rightarrow j
l^{+}l^{-} \nu_{l}\nu_{i}B_{H}B_{H}.
 \end{eqnarray}
In the above processes, we have assumed that the T-odd gauge bosons
$W_{H}$ and $Z_{H}$ mainly decay into $W B_{H}$ and $H B_{H}$,
respectively. For the $Higgs$ boson mass $M_{H}\leq120GeV$, its
dominant decay channel is $H \rightarrow b\overline{b}$ . In order
to ensure the cleanest event signature, only fully leptonic decay
modes of the gauge boson $W$ are considered. It is obvious that the
chain decay processes (9), (10), and (11) can lead to the $l^{\pm}$+
$jet$ + $E \hspace{-0.25cm}/$, $l^{\pm}$+ $jets$ + $E
\hspace{-0.25cm}/$, and $l^{+}l^{-}$+ $jet$ + $E \hspace{-0.25cm}/$
\hspace*{0.1cm} signatures. The production rates for these three
kinds of the signatures can be easily estimated by multiplying the
overall decay branching ratios to the effective production cross
sections for the above processes. For example, the production rate
of the $l^{\pm}$+ $jet$ + $E\hspace{-0.25cm}/$ signature can be
written as: $ \sigma_{B_{H}}\times B r(Q_{H} \rightarrow
jW_{H})\times B r(W_{H} \rightarrow WB_{H}) \times B r(W \rightarrow
l\nu_{l}) \approx \sigma_{B_{H}}\times0.57\times1\times0.32 \approx
0.18 \sigma_{B_{H}}$. The number of the raw signal events generated
per year at the $LHC$ are given in Fig.4, in which we have taken the
scale parameter $f=500GeV$ and the yearly integrated luminosity
$\pounds=100fb^{-1}$. One can see from this figure that there will
be several and up to hundreds of these kinds of the signal events to
be generated at the $LHC$ per year.

\begin{figure}[htb]
\begin{center}
\epsfig{file=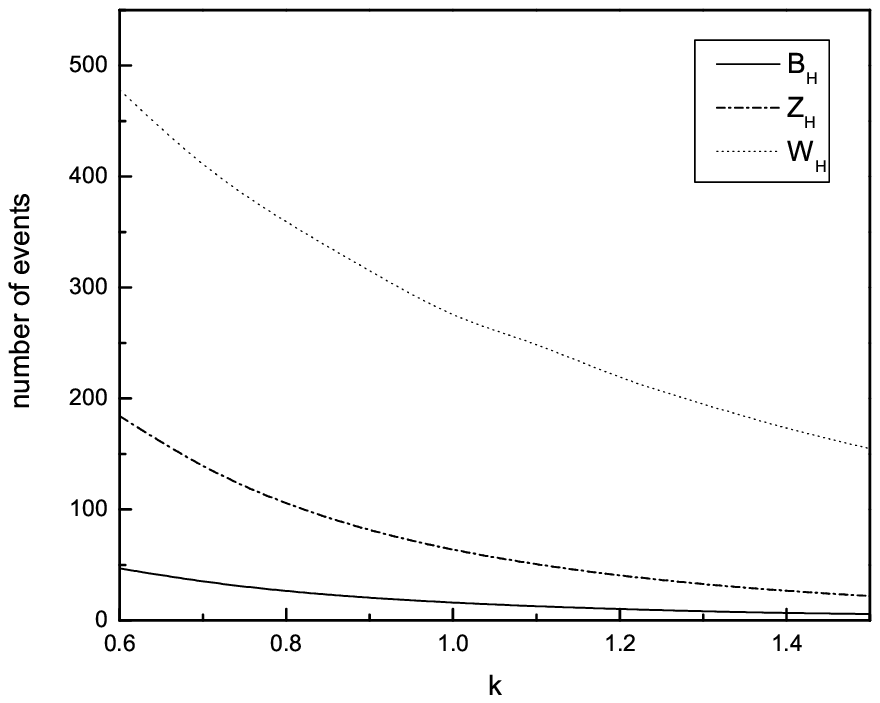,width=275pt,height=230pt}
 \vspace{-0.4cm}\caption{The number of the raw signal
 events versus the parameter $k$ for $f=500GeV$
\hspace*{1.8cm}
 and $\pounds=100fb^{-1}$.}
\end{center}
\end{figure}

In general, the photoproduction signals at the $LHC$ have two kinds
of backgrounds: irreducible and reducible backgrounds, which have
very similar final states or same final states as the signal, and
come from the photoproduction processes and the parton-parton
interaction processes, respectively. During the phase of the low
luminosity, one can use the large rapidity gap ($LRG$) way to
distinguish the photoproduction signal from the reducible
backgrounds [6,7,8,9]. At high luminosity (general about
$100fb^{-1}$) the reducible backgrounds can be suppressed by using
the dedicated very forward detectors ($VFDs$). Applying the
acceptance cuts, one can significantly suppress the irreducible
backgrounds coming from the $SM$ photoproduction processes, such as
$q\gamma\rightarrow jW$, $q\gamma\rightarrow q'WH$,
$q\gamma\rightarrow q'W^{\pm}W^{\mp}$, etc. To be certain, a
detailed simulation is needed, which has been extensively studied in
Refs.[7,8,9] and is beyond the scope of this paper.

\noindent{\bf 4. Photoproduction of the mirror quark associated with
a T-odd scalar}

The $LHT$ model predicts the existence of a complex T-odd triplet
scalar $\Phi$ with the mass in the range of $350GeV\sim1400GeV$ for
$m_{H}=120GeV$ and $500GeV\leq f \leq 2000GeV$. At the leading
order, its components $\phi^{\pm}$, $\phi^{0}$, and $\phi^{p}$ have
the same mass and can be produced via the parton-parton collision
processes at the $LHC$, which have been discussed in Ref.[27]. It
has been shown that the production cross sections are much small.
Although the T-odd scalar couples to ordinary quark and the mirror
quark at the order $v^{2}/f^{2}$ [15], to compare the
photoproduction with the partonic production of the T-odd scalars,
we consider photoproduction of the T-odd scalars in this section.
The  relevant Feynman diagrams are shown in Fig.5.
\begin{figure}[htb]
\begin{center}
\epsfig{file=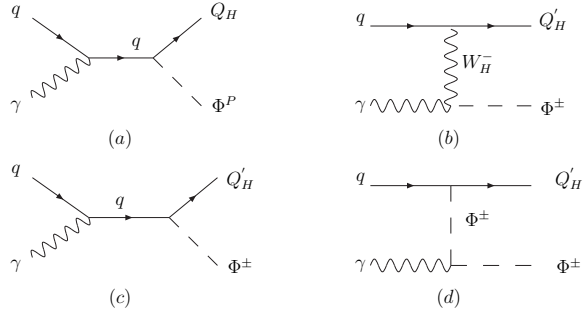,scale=0.6}
 \caption{Feynman diagrams for photoproduction of the mirror quark in association
\hspace*{2.0cm}
 with a T-odd scalar.}
 \label{ee}
\end{center}
\end{figure}

\vspace*{-0.8cm}

Using the relevant Feynman rules given in Ref.[15], the invariant
scattering amplitudes for photoproduction of the T-odd scalars can
be written as
\begin{eqnarray}
M_{\phi^{\pm}}=-\frac{e^{2}V_{ij}}{6\sqrt{2}S_{W}}\frac{\nu^{2}}{f}\bar{u}(P_{Q})
\{\frac{ie \epsilon \hspace{-0.20cm}/
 P_{L}}{S_{W}(\widehat{t}_{\phi}-M^{2}_{W_{H}})}+
\frac{MP_{L}}{2fM_{W_{H}}}[\frac{q^{i}(\not\!\!P_{q}+\not\!\!P_{\gamma})
\epsilon\hspace{-0.20cm}/}{2\hat{s}}
+\frac{P_{\gamma}^{\mu}\epsilon}{\widehat{t}_{\phi}-M_{\Phi}^{2}}]\}u(q),
\end{eqnarray}

\vspace*{-0.8cm}
\begin{eqnarray}
M_{\phi^{P}}=\frac{e^{2}q^{i}V_{ij}}{12\hat{s}}\frac{\nu^{2}}{f^{2}}
[\frac{1}{\sqrt{10}C_{W}M_{B_{H}}}+
\frac{1}{\sqrt{2}S_{W}M_{Z_{H}}}]M
\bar{u}(P_{Q})[P_{L}(\not\!\!P_{q}+\not\!\!P_{\gamma})\epsilon
\hspace{-0.20cm}/]u(P_{q}).
\end{eqnarray}
where $\widehat{t}_{\phi}=(P_{Q}-P_{q})^{2}$,
$\hat{s}=(P_{q}+P_{\gamma})^{2}$.

\begin{figure}[htb]
\begin{center}
\epsfig{file=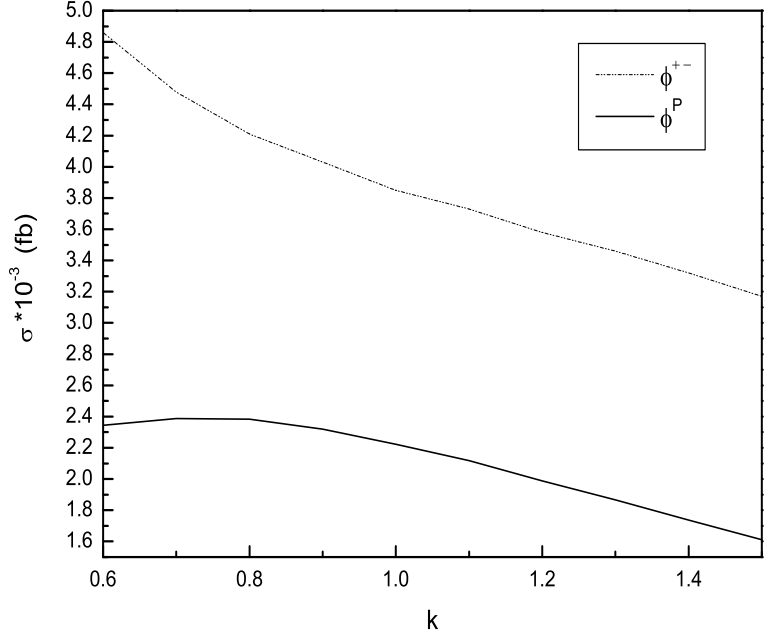,width=285pt,height=240pt}
 \vspace{-0.4cm}\caption{The production cross sections for the
 new  scalars as function of the coupling
\hspace*{2.0cm}
 constant $k$ for $f=500GeV$}
\end{center}
\end{figure}

Our numerical results are given in Fig.6, in which we have plotted
the production cross sections as functions of the coupling constant
$k$ for the scale parameter $f=500GeV$ and the other relevant free
parameters are taken to be same as section 3. One can see from this
figure that the production cross section is indeed much small. In
most of the parameter space of the $LHT$ model, the value of the
total cross section for photoproduction of the T-odd scalar in
association with the mirror quark is smaller than $0.01fb$. However,
compared that of partonic production of the T-odd scalar associated
the T-odd gauge boson, its background is also very small, which
mainly comes from the photon-induced processes.

In most of the parameter space of the $LHT$ model, the T-odd scalars
$\phi^{\pm}$ and $\phi^{p}$ mainly decay into $W^{\pm}B_{H}$ and
$HB_{H}$, respectively. For $m_{H}=120GeV$, $k=1$ and $f=1TeV$,
there are $Br(\phi^{\pm}\rightarrow W^{\pm}B_{H})\simeq 1$ and
 $Br(\phi^{p}\rightarrow HB_{H})\simeq 1$. Thus, photoproduction of
 the T-odd scalars at the $LHC$ can give the following final states:
\begin{eqnarray}
 q \gamma \rightarrow Q_{H}\phi^{p}\rightarrow j W_{H}
 H B_{H}\rightarrow j W H B_{H}B_{H} \rightarrow j  l  \nu_{l}
  b  \overline{b}  B_{H}B_{H},
\end{eqnarray}

\vspace*{-1.1cm}
\begin{eqnarray}
 q \gamma \rightarrow Q_{H}^{'}\phi^{\pm}\rightarrow j W_{H}^{\mp}
 W^{\pm} B_{H}\rightarrow j W^{\mp}W^{\pm}B_{H}B_{H} \rightarrow j
 l^{+}l^{-}\nu_{l}\overline{\nu}_{l} B_{H}B_{H},
\end{eqnarray}
which can induce the $l^{\pm}$+ $jets$ + $E \hspace{-0.25cm}/$ and
$l^{+}l^{-}$+ $jet$ + $E \hspace{-0.25cm}/$ signatures. The first
kind of signatures is same as that of $Eq.(10)$ and the second is
same as that of $Eq.(11)$. Thus, we have to say that it is more
difficult to detect the possible signatures of the mirror quark via
the subprocess $q\gamma\rightarrow Q_{H}\Phi$ than via the
subprosess $q\gamma\rightarrow Q_{H}^{'}W_{H}$.

\noindent{\bf 5. Pair production of the mirror quarks via $\gamma g$
and $\gamma\gamma$ collisions  }

At the $LHC$, the mirror quarks can be produced in pairs via
exchanging the T-odd gauge bosons ($B_{H}$ and $Z_{H}$) and gluon
exchange. It has been shown that, as long as their masses are not
too large, the mirror quarks can be copiously produced in pairs
[13]. However, the mirror quarks can also be produced in pairs via
$\gamma g$ and $\gamma\gamma$ collisions at the $LHC$. To completely
consider of the mirror quarks, we will consider their photon-induced
production in this section. The relevant Feynman diagrams are
displayed in Fig.7, in which Fig.7a is similar to that for the $SM$
process $g\gamma\rightarrow t\overline{t}$ and Fig.7b is similar to
that for the two-photon production of the supersymmetric pairs or
the top quark pairs.
\begin{figure}[htb]
\begin{center}
\epsfig{file=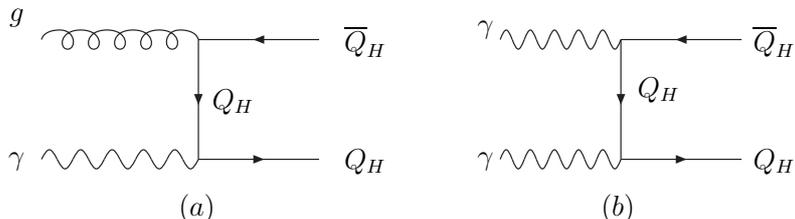,scale=0.9}
 \vspace{-0.8cm}
 \caption{Feynman diagrams for the photon-induced production
 of the mirror quark pair.}
 \label{ee}
\end{center}
\end{figure}

\vspace*{-0.6cm}

\begin{figure}[htb]
\begin{center}
\vspace{-0.5cm}
 \epsfig{file=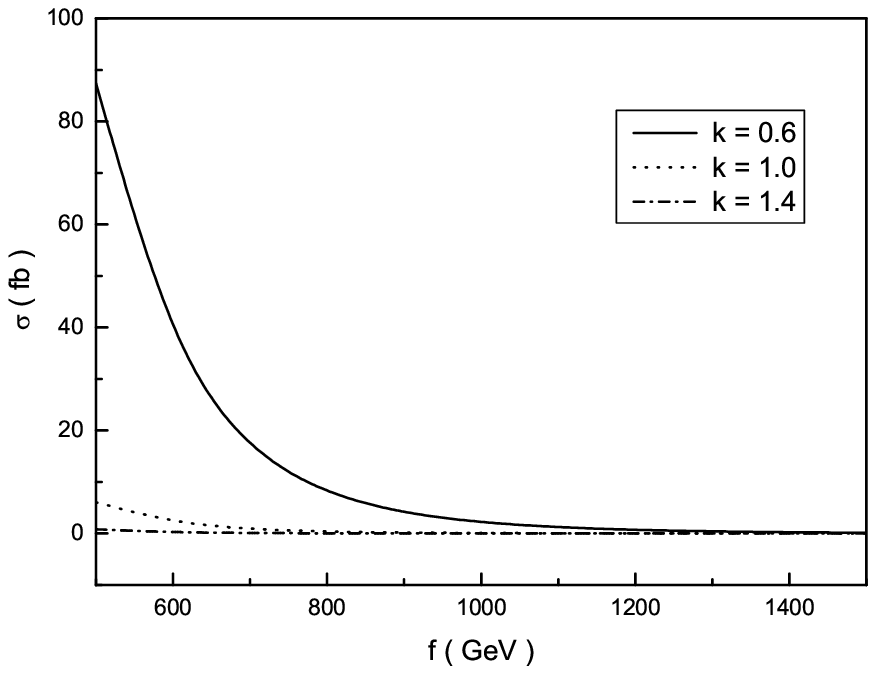,width=200pt,height=165pt}
\put(-115,-10){ (a)}\put(115,-10){ (b)}
 \hspace{0cm}\vspace{-0.25cm}
\epsfig{file=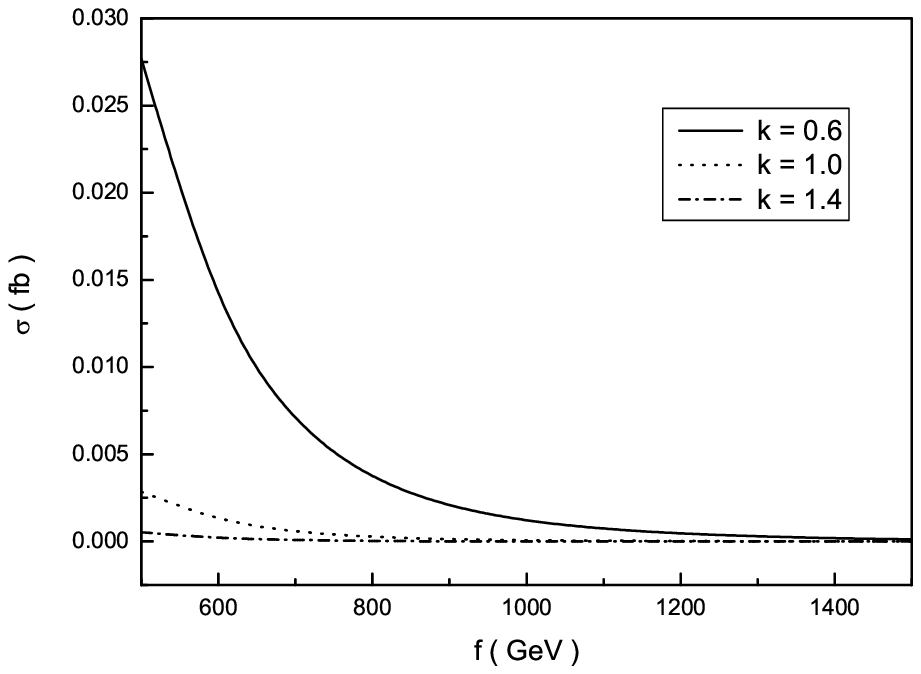,width=200pt,height=165pt} \hspace{-0.5cm}
 \hspace{10cm}\vspace{-1cm}
\vspace{0.5cm}
 \caption{The production cross sections induced by $\gamma g$
 ($a$) and $\gamma \gamma$ ($b$) collisions versus $f$
 \hspace*{1.8cm}
 value for different values of $k$.}
 \label{ee}
\end{center}
\end{figure}

\vspace*{-0.5cm}

It is obvious that the pair production cross sections of the mirror
quarks are only dependent on the scalar parameter $f$ and the
coupling parameter $k$. In Fig.8(a) and Fig.8(b) we present the
cross sections induced by $\gamma g$ and $\gamma\gamma$ collisions
versus $f$ value for different values of the parameter $k$. One can
see from these figures that the photon-induced production of the
mirror quark pairs are mainly generated by the subprocess $g
\gamma\rightarrow \overline{Q}_{H}Q_{H}$ and the effective cross
section $\sigma_{g\gamma}$ is larger than $\sigma_{\gamma\gamma}$ at
least by three orders of magnitude. Certainly, this is because the
gluon luminosity is much larger than the photon luminosity and the
coupling of the gluon to the mirror quarks is stronger than that for
the photon. For $k=0.6$ and $500GeV\leq f\leq1500GeV$, the values of
the cross section $\sigma_{g\gamma}$ is in the range of
$2.1\times10^{-3}fb\sim87.5fb$.

If we assume that the mirror quark decays into $Wj$($j$=$u$, $c$,
$d$, or $s$) and focus our attention only on the pure leptonic decay
modes for the $SM$ gauge boson $W$, then pair production of the
mirror quark can induce the $l^{+}l^{-}$+ $jets$ + $E
\hspace{-0.25cm}/$ signature. The total number of this kind of
signal are given in Fig.9 for $f=500GeV$, in which we have taken the
integral luminosity $\pounds=100fb^{-1}$ and $Br(Q_{H}\rightarrow
W_{H}j)\simeq 57$\%. One can see from this figure that there will be
several and up to hundreds of the $l^{+}l^{-}$+ $jets$ + $E
\hspace{-0.25cm}/$ signal events to be generated at the $LHC$ per
year.

\begin{figure}[htb]
\begin{center}
\epsfig{file=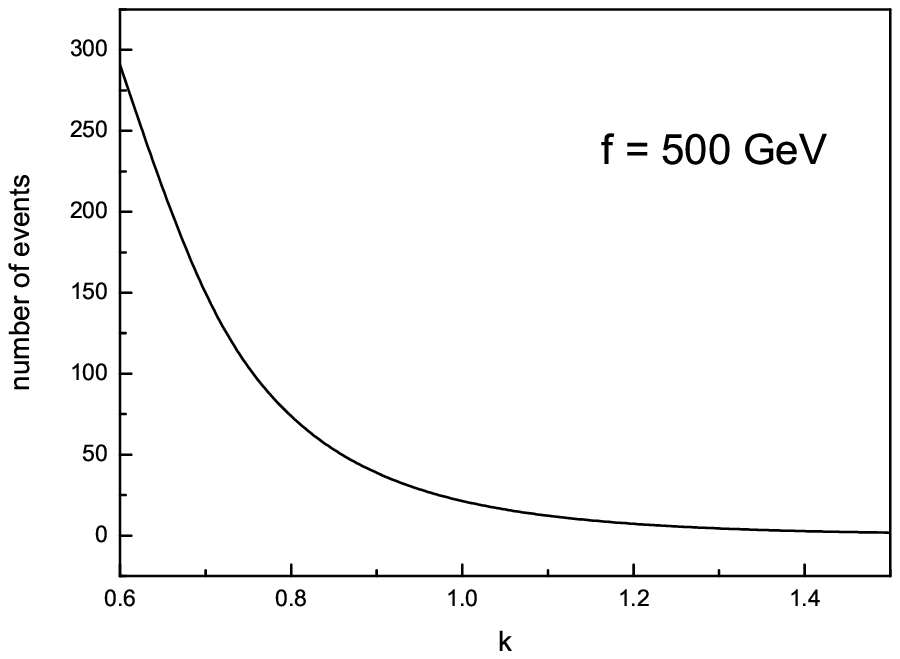,width=285pt,height=240pt}
 \vspace{-0.8cm}
 \caption{The number of the $l^{+}l^{-}$+ $jet$ + $E
 \hspace{-0.25cm}/$ signature events for $f=500GeV$ and
\hspace*{1.8cm}
  $\pounds=100fb^{-1}$.  }
 \label{ee}
\end{center}
\end{figure}

The reducible $SM$ backgrounds of pair production of the mirror
quark via $\gamma p$ and $\gamma\gamma$ interactions mainly come
from $t\overline{t}$, which both top quarks decay leptonically, and
$W^{+}W^{-}jj$, which the two jets originate from initial-state
radiation. Although the reducible background is several orders of
magnitude larger than the signal, one expects that it can be reduced
to the same level as the irreducible background by using $LRG$
condition and the dedicated $VFDs$ [6,7,8,9]. The irreducible
backgrounds mainly come from the photon-induced processes $\gamma g
\rightarrow t\overline{t}$ and $\gamma\gamma\rightarrow
t\overline{t}$. The ratio of the signal over the square root of the
background $R=S/\sqrt{B}$ (called the statistical significance ) is
given in Fig.10 in which we have taken $f=500GeV$ and
$\pounds=100fb^{-1}$. One can see from this figure that, with
reasonable values of the free parameters, the values of $R$ can be
significantly large.
\begin{figure}[htb]
\begin{center}
\epsfig{file=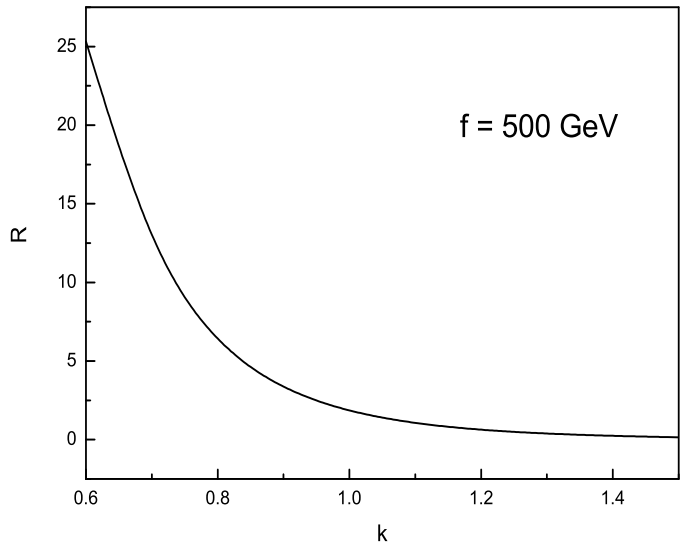,width=285pt,height=240pt}
 \vspace{-0.9cm}
 \caption{The statistical significance $R$ as a function
  of the parameter $k$ for $f=500GeV$.}
 \label{ee}
\end{center}
\end{figure}

  The numerical results of Fig.10 are obtained in the case of no
any cut applied. Similar with Refs.[7,8,9], if we apply acceptance
cuts on the final state particles, the irreducible backgrounds can
be significantly suppressed and the value of R should be enhanced.
Thus, the possible signals of the mirror quarks might be detected
via $g\gamma$ collision at the $LHC$.

\noindent{\bf 6. Conclusions and discussions}

The photon-induced processes at the $LHC$ provide clean experimental
conditions due to absence of the proton remnants. Well defined final
states can be easily selected and precisely reconstructed.  To some
extend, the $LHC$ can be considered as a high-energy $\gamma-\gamma$
or $\gamma-p$ collider, which offer interesting possibilities for
studying the electroweak sector and for searching new physics up to
$TeV$ scale. One expects that the $\gamma-\gamma$ or $\gamma-p$
collision at the $LHC$ should give complementary and interesting
results for the tests of the $SM$ and for searching of new physics.
Thus, considering production of new particles via  $\gamma-\gamma$
or $\gamma-p$ collision at the $LHC$ is very interesting. It will be
helpful to detect the possible signatures of new physics models at
the $LHC$.

The $LHT$ model is one of the attractive little $Higgs$ models that
is not only consistent with electroweak precision tests but also
predicts the existence of the heavy T-odd $SU(2)$ doublet fermions,
which are called the mirror fermions of the $SM$ fermions. These new
particles might produce the observability signatures in future high
energy collider experiments. In this paper, we consider the
photon-induced production of the first and second generation mirror
quarks and further discuss its possible signatures at the $LHC$.

The effective production cross sections of the mirror quarks at the
$LHC$ via the subprocesses $q \gamma \rightarrow Q_{H}B_{H}(Z_{H})$,
$q \gamma \rightarrow Q_{H}^{'}W_{H}^{\pm}$, $q \gamma \rightarrow
Q_{H}^{'}\phi^{\pm}$, $q \gamma \rightarrow Q_{H}\phi^{p}$, $g
\gamma \rightarrow \overline{Q}_{H}Q_{H}$, and $\gamma\gamma
\rightarrow \overline{Q}_{H}Q_{H}$ are calculated. Our numerical
results show that the values of cross sections for all of these
production channels are strongly dependent on the scale parameter
$f$ and the coupling parameter $k$. Their values decrease quickly
 as $f$ and $k$
increase. However, as long as the mirror quark is not too heavy,
i.e. the parameters $f$ and $k$ are not too large, it can be
significantly produced via some of these processes at the $LHC$. For
example, for $0.6\leq k \leq 1$ and $500GeV\leq f\leq1500GeV$, the
cross section value of the subprocess $q \gamma \rightarrow
Q_{H}^{'}W_{H}$ is larger than $0.1fb$ and can reach $82fb$, and the
value of the cross section for the subprocess $q g \rightarrow
\overline{Q}_{H}Q_{H}$ is in the range of $2.1\times 10^{-3}fb \sim
88fb$.

The different chain decay channels of the mirror quark can give
different experimental signatures. The possible signatures of the
mirror quarks generated from the partonic processes have been
studied in Ref.[13]. Considering the domaint decay channels of the
mirror quarks, the possible signatures generated by the
photon-induced processes are also discussed in this paper. We find
that some of the photon-induced processes can produce the same
signals as those for the partonic processes. However,  because of
the clean experimental conditions and the well defined final states,
they might be easily detected. For the photon-induced signatures,
most of the backgrounds coming from the partonic interactions,
called reducible background, can be significantly omitted by using
the $LRG$ technique and the dedicated $VFDs$. The irreducible
background generated by $\gamma-\gamma$ or $\gamma-p$ interaction
can be largely suppressed by applying acceptance cuts. Certainly, it
should be further studied.

All of our numerical results are obtained in the case of only
considering the elastic photon contributions. If the contributions
of the inelastic photons are included, the corresponding cross
sections are increased by about a factor of three [2]. It is obvious
that the irreducible backgrounds
 are also increased. Furthermore, the strong interactions between
protons, the so called rescatting  effects, can suppress the
photon-induced cross sections, which depends on the invariant mass
of the exclusively produced state, such as $Q_{H}W_{H}$ or
$\overline{Q}_{H}Q_{H}$. This kind of correction effects is ignored
in our numerical results. In our simply phenomenology analysis, we
have taken the T-odd gauge boson $B_{H}$ as missing energy. If we
assume the the T-parity is violated, then $B_{H}$ can decay into
$WW$ and $ZZ$ pairs, which can induce different signatures.

\noindent{\bf Acknowledgments}

This work was supported in part by the National Natural Science
Foundation of China under Grants No.10675057, the Natural Science
Foundation of the Liaoning Scientific Committee(No.20082148), and
Foundation of Liaoning  Educational Committee(No.2007T086).

\end{document}